\documentclass[%
 reprint,
 amsmath,amssymb,
 aps,
]{revtex4-2}
\usepackage{amsmath,amsfonts,latexsym,amssymb,graphics,epsfig,subfigure,color,makeidx}
\usepackage{lipsum}
\usepackage{float}
\usepackage[dvipsnames]{xcolor}
\usepackage[colorlinks,
            linkcolor=blue,
            anchorcolor=blue,
            citecolor=green
            ]{hyperref}
\usepackage{tikz}
\usetikzlibrary{positioning}
\usetikzlibrary{patterns}
\usepackage{booktabs}  % for \toprule, \midrule, \bottomrule
\usepackage{makecell}
\usepackage{graphicx}% Include figure files
\usepackage{dcolumn}% Align table columns on decimal point
\usepackage{bm}% bold math
\usepackage{acro}

\DeclareAcronym{tovf}{
  short = TOVF,
  long = TOV Firehole,
  short-plural = s,
  long-plural = s
}

\DeclareAcronym{tovfo}{
  short = TOVFO,
  long = TOVF Open,
  short-plural = s,
  long-plural = s
}

\DeclareAcronym{tovfb}{
  short = TOVFB,
  long = TOVF Boxed,
  short-plural = s,
  long-plural = s
}

\begin{document}

\def\newacronym#1#2#3{\gdef#1{#3 (#2)\gdef#1{#2}}}
\newacronym{\tov}{TOV}{Tolman–Oppenheimer–Volkoff}
\newacronym{\eos}{EOS}{equation of state}
\newacronym{\bh}{BH}{black hole}
\newacronym{\isco}{ISCO}{innermost stable circular orbit}
\newcommand{\PL}[1]{{\textcolor{red}{\sf{[PL: #1]}}}}
\newif\ifshowrev \showrevtrue
\newcommand{\rev}[1]{\ifshowrev{\color{blue}#1}\else#1\fi}
\newcommand{\CL}[1]{{\textcolor{red}{\sf{[CHECK: #1]}}}}
\def\icore{\emph{interior core}}
\def\tshell{\emph{transition layer}}
\def\splateau{\emph{Schwarzschild-like plateau}}
\def\atail{\emph{self-similar asymptotic tail}}
\def\lr{\emph{light ring}}
\def\lrs{\emph{light rings}}

%\graphicspath{{Images/}}

\title{On the Stability of Tolman–Oppenheimer–Volkoff Fireholes}

\author{Pablo Laguna$^1$, Octavio Guerrero$^1$, Kostas Kokkotas$^2$}
\affiliation{
\small$^{1}$Center for Gravitational Physics, Weinberg Institute, Department of Physics, The University of Texas at Austin, Austin, TX 78712, USA\\
\small$^{2}$Theoretical Astrophysics, Institute for Astronomy and Astrophysics, University of T\"ubingen, 72076 T\"ubingen, Germany}

\begin{abstract}
We investigate the radial stability of singular
Tolman--Oppenheimer--Volkoff fireholes with a
radiation equation of state. These configurations
contain a naked, negative point-mass singularity
surrounded by a positive-density fluid, whose mass profile develops a Schwarzschild-like plateau before
approaching an extended, self-similar atmosphere.
Using the Pr\"ufer phase method to identify the
fundamental radial mode, we consider two outer
boundary conditions: vanishing Lagrangian pressure
perturbation (open firehole) and vanishing radial displacement
(boxed firehole). For both, the fundamental squared frequency
changes from positive to negative as the outer
boundary exceeds a critical radius $R_c$. We find
that $R_c$ follows approximately the same scaling
with the central mass parameter as the equilibrium
firehole radius $R$. However, the stability thresholds
differ significantly: $R_c^{\rm O}<R$ for the open firehole, whereas
$R_c^{\rm B}\simeq R$ for the boxed one.
The regular zero mode is the derivative of the equilibrium
family with respect to the inner lapse normalization at
fixed central mass. This gives the boxed threshold a
turning-point interpretation and, together with a scaling
symmetry of the TOV equations, implies $R_c^{\rm B}\to R$
in the limit of large central mass. Variational bounds show
that the rigid wall is the most stabilizing separated,
self-adjoint outer boundary condition and that the unbounded configuration is
unstable. Tolman--Oppenheimer--Volkoff fireholes are
therefore radially unstable unless confined within
approximately their firehole radius.
\end{abstract}

\maketitle

\section{Introduction}
\label{sec:introduction}

The Tolman--Oppenheimer--Volkoff (TOV) equations provide
the fundamental description of static, spherically
symmetric self-gravitating fluids in general relativity
\cite{Tolman1939,OppenheimerVolkoff1939}.
For a specified equation of state, their solutions
determine the equilibrium structure of relativistic
stars and the relation between their gravitational mass,
radius, and internal matter distribution. They have
consequently played a central role in relativistic
stellar astrophysics, most prominently in models of
neutron stars, but also in relativistic treatments of
white dwarfs and other compact stellar configurations.
Much of our understanding of the maximum masses,
compactness, and stability of relativistic stars
ultimately rests on properties of solutions of the
TOV equations.

The equilibrium solutions also provide the starting
point for studying the response and dynamics of
relativistic stars. Radial perturbations lead to the
relativistic stellar pulsation equations developed by
Chandrasekhar \cite{Chandrasekhar1964ApJ,
Chandrasekhar1964PRL}, whose eigenvalues determine the
onset of radial instability. Nonradial perturbations
characterize stellar oscillations and tidal
deformability, including the tidal Love numbers and
the approximate universal I--Love--Q relations
\cite{Hinderer2008,YagiYunes2013}. At the fully
nonlinear level, TOV solutions and their generalizations
also form the building blocks for initial data and
numerical-relativity simulations of isolated and binary
compact stars \cite{Cook2000,BaumgarteShapiro2003,
Duez2010}.

Nearly all these applications, however, concern the
regular branch of TOV solutions, for which the density
and pressure remain finite at the origin and the
enclosed mass vanishes as $r\to0$. Much less attention
has been given to singular solutions. A particularly
intriguing example was identified by Zurek and Page
\cite{ZurekPage1984}, and later considered by
't~Hooft \cite{tHooft1998}, in studies of
self-gravitating thermal radiation. Integrating the
TOV equations inward from a radiation atmosphere,
Zurek and Page found solutions with positive fluid
density everywhere but no event horizon. Instead,
these configurations develop a high-density region
near the nominal Schwarzschild radius and approach,
at small radii, a geometry dominated by a negative
point mass. Their mass profiles exhibit an
approximately Schwarzschild-like plateau, suggesting
a compact central object surrounded by an extended
radiation atmosphere. Zurek and Page emphasized the
thermodynamic properties of these configurations,
including the possibility that the strongly
blueshifted radiation carries an entropy comparable
to the Bekenstein--Hawking entropy.

Despite these remarkable properties, the dynamical
stability of the singular branch, which we refer to
as TOV fireholes (TOVFs), remains largely unexplored.
Anastopoulos and Savvidou~\cite{anastopoulos} classified
singular TOV solutions, showed that near the center they
are locally isomorphic to negative-mass Schwarzschild,
argued that the strongly repulsive singularity stabilizes
the surrounding matter, and undertook a preliminary
investigation of their dynamical stability.
Here we show that, despite the repulsive core, TOVFs
are radially unstable unless the fluid is confined
within approximately the firehole radius.
The existence of a static equilibrium solution does
not establish its stability under perturbations.
Moreover, unlike ordinary relativistic stars, TOVFs
with a radiation equation of state have no finite
radius at which the pressure vanishes. Their fluid
atmosphere extends indefinitely and approaches a
self-similar, non-asymptotically flat configuration.
Consequently, the radial pulsation problem requires
particular care in specifying the inner regularity
condition and the outer boundary condition.
An important question is whether the compact firehole
can support stable radial oscillations independently
of its extended atmosphere, or whether stability
requires a mechanism that confines the surrounding
fluid.

In this work, we investigate the radial stability
of TOVFs with a radiation equation of state,
$p=\rho/3$, using Chandrasekhar's relativistic
pulsation equations. We consider two boundary-value
problems at a prescribed outer radius $R_{\rm out}$:
vanishing Lagrangian pressure perturbation,
which we call TOVF Open (TOVFO), and vanishing
radial displacement at a rigid boundary,
which we call TOVF Boxed (TOVFB). Using the
Pr\"ufer phase method, we identify the fundamental
radial mode and determine the critical radius
$R_c$ at which its squared frequency changes
sign. We find that the critical radii approximately
follow the same scaling with the central negative
mass parameter as the characteristic firehole radius
$R$ and plateau mass $M$.

Our principal result is that the stability threshold
depends significantly on the outer boundary condition.
For TOVFO, the fundamental mode becomes unstable
at a critical radius $R_c^{\rm O}<R$, whereas for
TOVFB, the threshold nearly coincides with the
firehole radius, $R_c^{\rm B}\simeq R$.
A variational analysis establishes the stabilizing
effect of the rigid boundary and explains the ordering
of these critical radii. We show that the regular
zero-frequency mode is the displacement connecting
neighboring equilibria, which explains why
$R_c^{\rm B}$ approaches $R$. Since even the boxed
configuration, which is the most stabilizing
separated, self-adjoint outer boundary condition, develops a radial
instability when the fluid domain extends beyond
approximately the firehole radius, TOVFs cannot be
regarded as radially stable isolated objects without
an additional mechanism that confines or otherwise
stabilizes their extended atmosphere. A variational
argument extends this conclusion to the complete,
unbounded atmosphere.

\acuse{tovf,tovfo,tovfb}The paper is organized as follows. In Sec.~\ref{sec:equilibria},
we construct the TOVF equilibrium solutions,
describe their interior and exterior structure,
and examine the scaling relations governing
their characteristic mass and radius.
In Sec.~\ref{sec:4}, we formulate the radial pulsation
problem, determine the stability thresholds for
open and boxed boundary conditions, investigate
their eigenfunctions and variational properties,
relate the zero modes to neighboring equilibria,
and derive bounds valid for general confinement
and for the unbounded atmosphere.
We summarize our results and discuss their
implications in Sec.~\ref{sec:conclusions}.

\begin{figure}[htbp]  % h = "here" (tries to place image at this point)
    \centering
    \includegraphics[width=\linewidth]{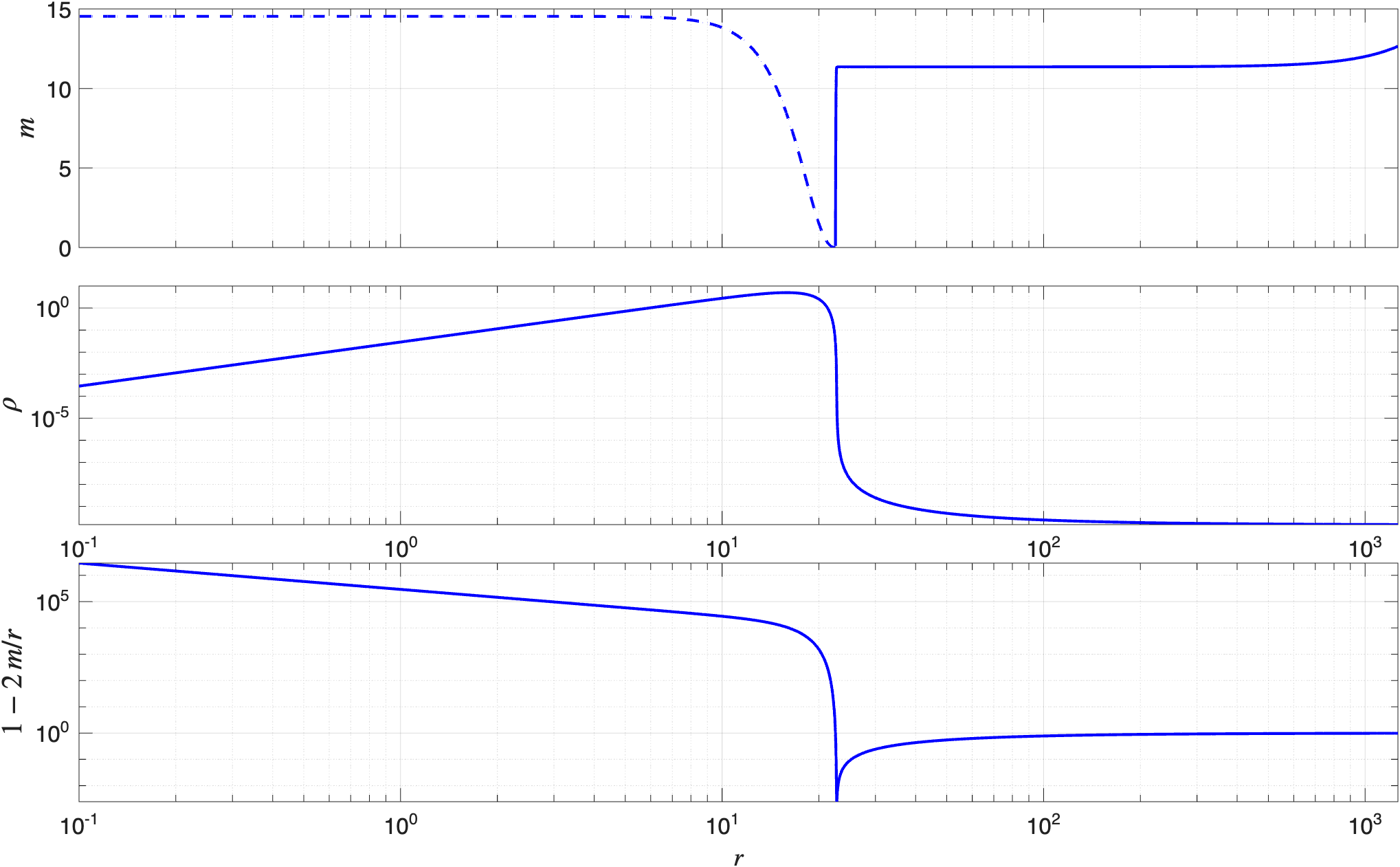}  % Adjust width
    \caption{Zurek--Page singular solution to the TOV equations, as in Fig.~2 of Ref.~\cite{ZurekPage1984}. The outer radius was at $1,265$ with $m = 12.65$ and  $\rho = 1.5\times 10^{-10}$. From top to bottom are the  mass function $m$, the density $\rho$, and the quantity $1-2\,m/r$. The dashed line in the top panel was obtained from $-m(r)\times 10^{-4}$.}
    \label{fig:ZP_Fig2}
\end{figure}

\section{TOV Fireholes}
\label{sec:equilibria}

A static, spherically symmetric spacetime metric may be written as
\begin{equation}
ds^2 = -e^{2\nu(r)}dt^2 + e^{2\lambda(r)}dr^2 + r^2 \,d\Omega^2\,,
\label{eq:metric}
\end{equation}
where $d\Omega^2 = d\theta^2 + \sin^2{\theta}\, d\phi^2$. If the matter is a perfect fluid, the Einstein equations, in $G=c=1$ units, yield the so-called TOV equations~\cite{OppenheimerVolkoff1939, Tolman1934-TOLRTA}
\begin{eqnarray}
        m' &=& 4\,\pi\, r^2 \rho,  \label{eq:TOV1}\\
        p' &=& - \frac{ ( \rho + p) (m + 4\,\pi\, p\, r^3)}{r (r - 2\, m)},\label{eq:TOV2}\\
        \nu' &=& -\frac{p'}{(\rho+p)}\,, \label{eq:TOV3}
\end{eqnarray}
where primes denote derivatives with respect to the radial coordinate and $e^{-2\,\lambda} \equiv 1-2\,m/r$. In these equations, $\rho$ is the total mass-energy density, $p$ is the pressure, and $m$ is the effective mass enclosed within the radius $r$.

As in the Zurek and Page~\cite{ZurekPage1984} study, we will focus on the $\Gamma$-law \eos{} $p = (\Gamma -1)\,\rho$ with $\Gamma = 4/3$. Fig.~\ref{fig:ZP_Fig2} reproduces Fig.~2 of Ref.~\cite {ZurekPage1984}. The integration of the \tov{} equations started at a radius $1,265$ with $m = 12.65$ and  $\rho = 1.5\times 10^{-10}$ and proceeded inwards. From the behavior of the mass function $m$ in the top panel, we observe four domains: an \icore{} where the mass function is approximately constant, with $m \simeq -\mu < 0$; a \tshell{} where $m$ grows rapidly when the density reaches a peak; and a \splateau{} where $m \simeq M$ that ends with a linear \atail{.} It is important to emphasize that with a $\Gamma$-law \eos{}, the density, and thus the pressure, never vanishes at a finite radius. There is no surface radius. This issue will play an essential role in the radial stability analysis. Furthermore, with the metric of the form in Eq.~(\ref{eq:metric}), the presence of a horizon is signaled by the location where $ 1-2\,m/r$ vanishes. This quantity is displayed in the bottom panel of Fig.~\ref{fig:ZP_Fig2}. For the singular solutions considered here and the one in Fig.~\ref{fig:ZP_Fig2}, the quantity $1-2m(r)/r$ remains positive: the negative-mass singularity is not enclosed by a horizon.

With $\Gamma = 4/3$, Eq.~(\ref{eq:TOV3}) yields $-4\,\nu' = (\ln{\rho})'$. Thus, $4\,\pi\,\rho \,e^{4\,\nu} = K$, with $K$ an integration constant. We introduce the definition $y \equiv e^{2\,\nu}$, so the \tov{} equations become
\begin{eqnarray}
        m' &=& K\left(\frac{r}{y}\right)^2 ,  \label{eq:tov1a}\\
        y' &=& \left(\frac{y}{r}\right)\left[\frac{2\,m}{r} + \frac{2\,K}{3}\,\left(\frac{r}{y}\right)^2\right]\left(1 - \frac{2 \,m}{r}\right)^{-1}\,.\label{eq:tov2a}
\end{eqnarray}
Without loss of generality, we set $K=1$, which fixes the overall scaling associated with the first integral of Eq.~(\ref{eq:TOV3}). The singular solutions nevertheless retain a second integration constant, denoted below by $A$, which fixes the normalization of the inner lapse and must be specified when defining a one-parameter family. Other solutions can be obtained with the appropriate rescaling.

Close to the origin, in the \icore{}, Eq.~(\ref{eq:tov2a}) reads
$y' \simeq -y/r$, and with Eq.~(\ref{eq:tov1a}), we get that
\begin{eqnarray}
    y &\simeq& \frac{A}{r} \label{eq:ycore}\\
    m &\simeq& -\mu +\frac{r^5}{5\,A^2}\label{eq:mcore}\,,
\end{eqnarray}
where $A>0$ is the remaining integration constant. The second expression can be rewritten as
\begin{eqnarray}
    \frac{m}{\mu} &\simeq& -1 +\left(\frac{r}{R_*}\right)^5\label{eq:mcore2}\,,
\end{eqnarray}
where $R_*^5 \equiv 5\,\mu\,A^2$. This scale is defined by extrapolating the inner expansion.
It need not coincide with the numerically determined radius
where $m(r)=0$. $R_*$ is the natural scale at which the positive fluid mass becomes comparable to the negative point mass. It also shows that the inner expansion depends on $r$ through $r/R_*$.

The \tshell{} is characterized by the rapid
growth of the mass function and by a minimum of
$e^{-2\lambda}=1-2m/r$, after which the mass function
approaches a \splateau{}. We define the
characteristic firehole radius $R$ as the location of
this minimum, or equivalently, the maximum of
$e^{2\lambda}$. The corresponding mass $M=m(R)$
defines the firehole mass. Unlike the surface of an
ordinary star, $R$ is not associated with vanishing
pressure, but is determined entirely by the
equilibrium geometry. By construction,
$
\left.\left(m/r\right)'\right|_{R}=0,
$
which implies
$m'(R)=M/R$, and
$4\pi R^3\rho(R)=M$.
This geometrical definition of $R$ provides
a natural reference scale for the critical
radii obtained from the radial stability
analysis.

From Fig.~\ref{fig:ZP_Fig2}, at large radius, the numerical solutions approach a linear \atail{}, suggesting $m=a\,r$. This requires from Eq.~(\ref{eq:tov1a}) that $\,(r/y)^2 = a$, and hence Eq.~(\ref{eq:tov2a}) yields
\begin{equation}
   1 = \frac{2\,a + 2\,a/3}{1 - 2 \,a}  \,.
\end{equation}
Thus, $a = 3/14$. Therefore, the asymptotic solutions are
\begin{eqnarray}
y &\simeq& \sqrt{\frac{14}{3}}\,r\label{eq:nu3}\\
m &\simeq& \frac{3}{14}\,r \label{eq:m3} \,.
\end{eqnarray}
Thus, the solutions approach a common
self-similar asymptotic branch. Furthermore, since $y=e^{2\nu}$ grows linearly
with $r$, these spacetimes are not asymptotically flat.

\begin{figure}[htbp]  % h = "here" (tries to place image at this point)
    \centering
    \includegraphics[width=\linewidth]{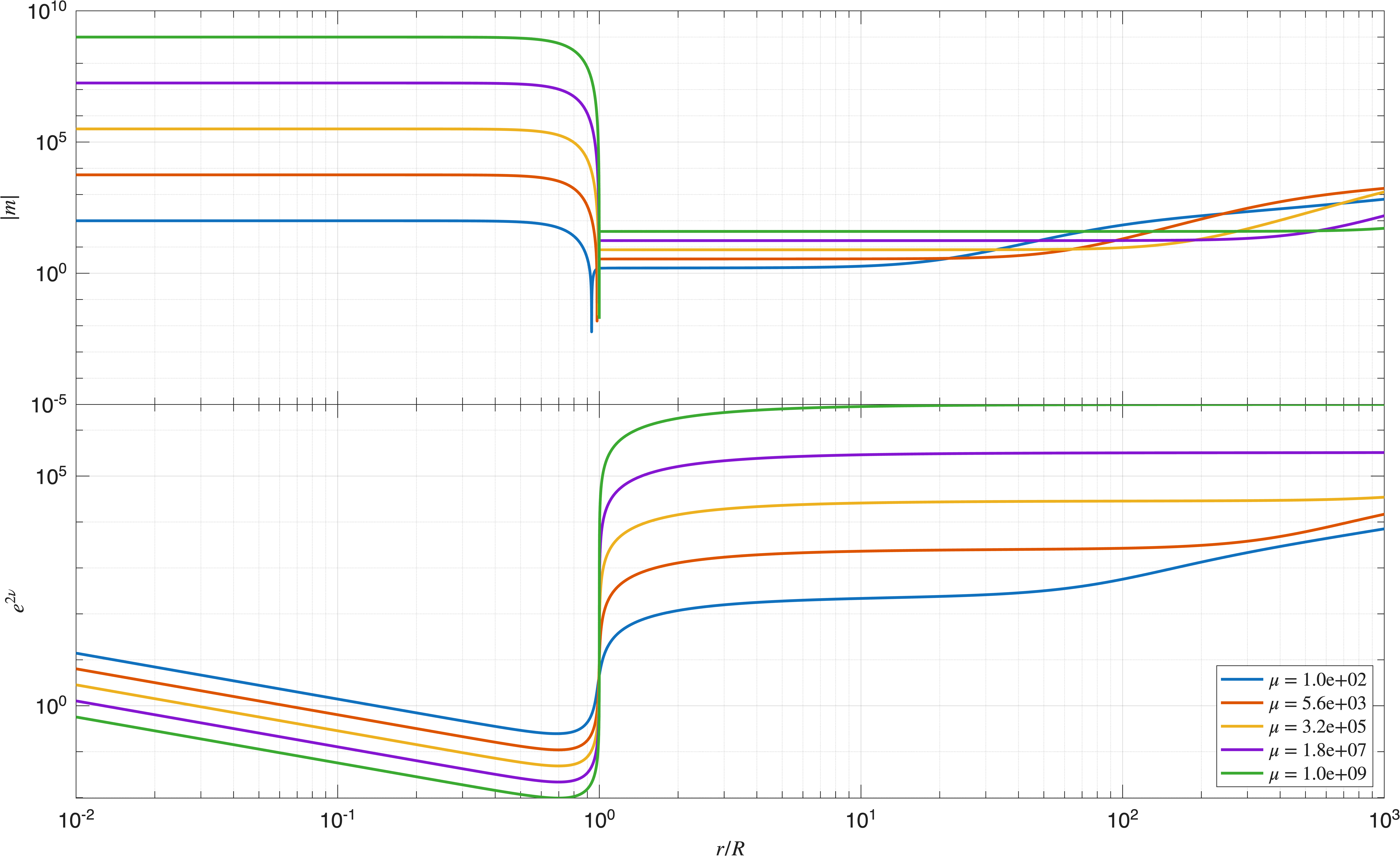}  % Adjust width
    \caption{Solutions in the range $10^2 \le \mu \le 10^9$, using as initial conditions Eqs.~(\ref{eq:ycore}) and (\ref{eq:mcore}) with $q=0$. The top panel shows $|m|$ and the bottom panel $y = e^{2\,\nu}$, as functions of $r/R$.}
    \label{fig:Fig1a}
\end{figure}

\begin{figure}[htbp]  % h = "here" (tries to place image at this point)
    \centering
    \includegraphics[width=\linewidth]{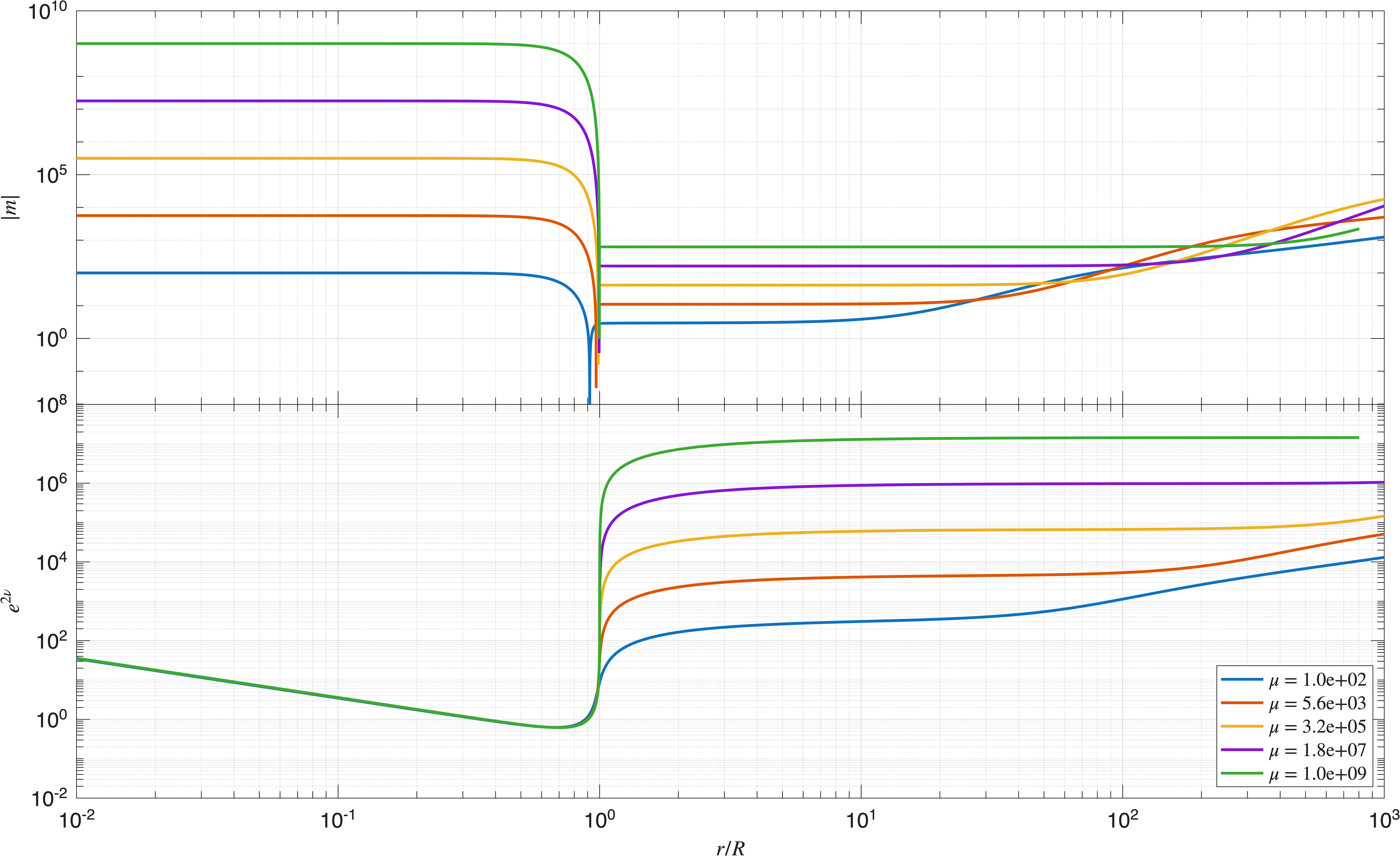}  % Adjust width
    \caption{Similar to Fig.~\ref{fig:Fig1a} but with $q=1/3$.}
    \label{fig:Fig1b}
\end{figure}

We parametrize  the second integration
constant $A$, associated with the normalization of
the inner lapse, as
\begin{equation}
A=\frac{\mu^q}{\sqrt{5}},
\label{eq:A_q}
\end{equation}
where the numerical factor is introduced for
convenience.
Fig.~\ref{fig:Fig1a} shows solutions for the case $q=0$. The top panel shows $|m|$ and the bottom panel $y = e^{2\,\nu}$ as functions of $r/R$.  Fig.~\ref{fig:Fig1b} shows the corresponding solutions for the case $q=1/3$. In both families, the mass function exhibits the same qualitative structure: a negative mass
interior core, a rapid transition layer, an
approximately \splateau{}, and
an extended self-similar atmosphere. The principal difference between the families is
the scaling of the characteristic length
and mass scales with $\mu$.

An enlightening form of displaying results is by following 't~Hooft~\cite{tHooft1998} and introducing the variables
\begin{eqnarray}
    Z &\equiv& e^{2\,\lambda} = \left(1-\frac{2\,m}{r}\right)^{-1} \\
    X &\equiv& \frac{e^{2\,\nu}}{r} = \frac{y}{r}\,.
\end{eqnarray}

\begin{figure}[htbp]  % h = "here" (tries to place image at this point)
    \centering
    \includegraphics[width=\linewidth]{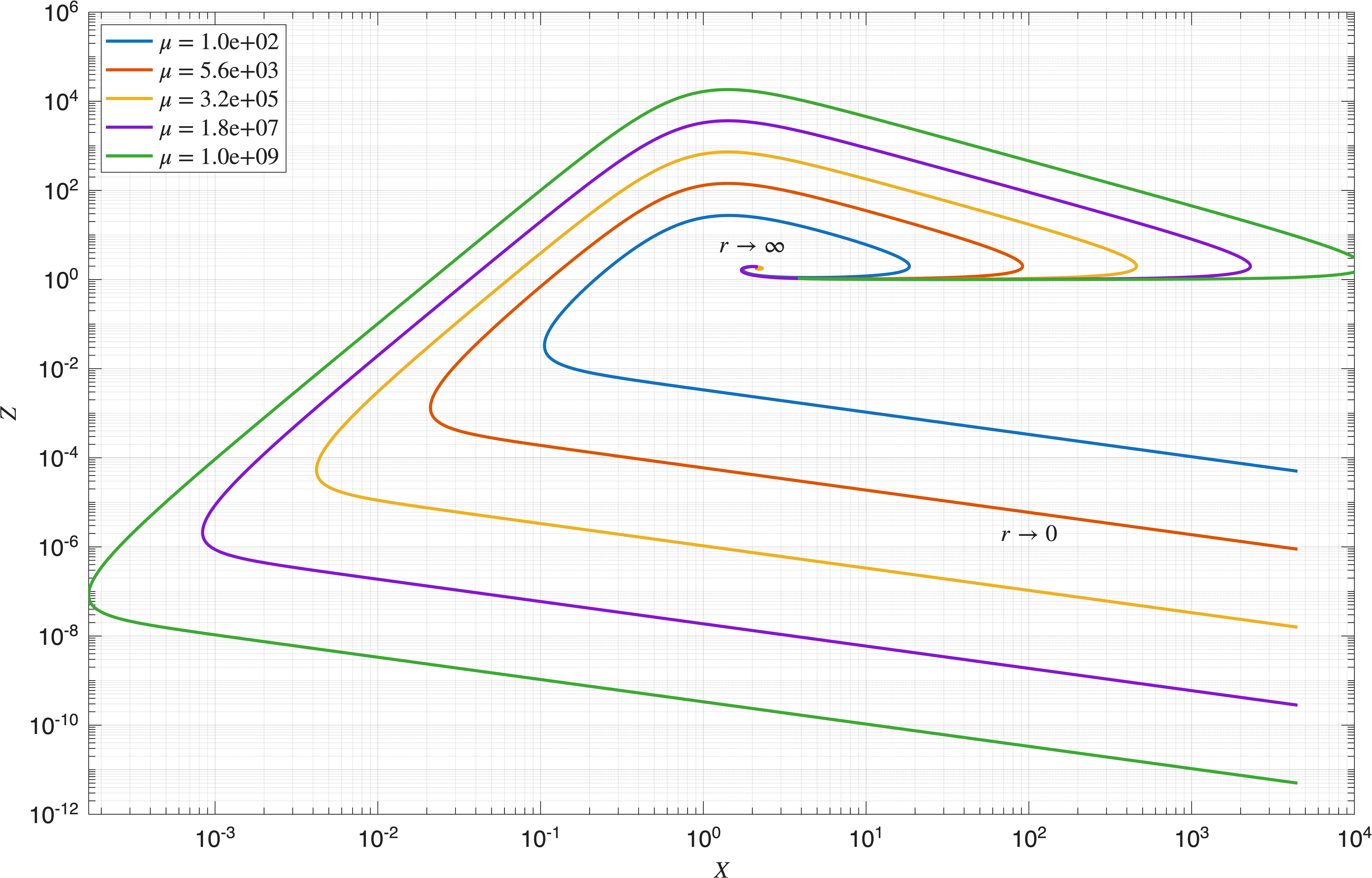}  % Adjust width
    \caption{Singular solutions from Fig.~\ref{fig:Fig1a} in terms of variables $Z$ and $X$.}
    \label{fig:Fig2a}
\end{figure}

\begin{figure}[htbp]  % h = "here" (tries to place image at this point)
    \centering
    \includegraphics[width=\linewidth]{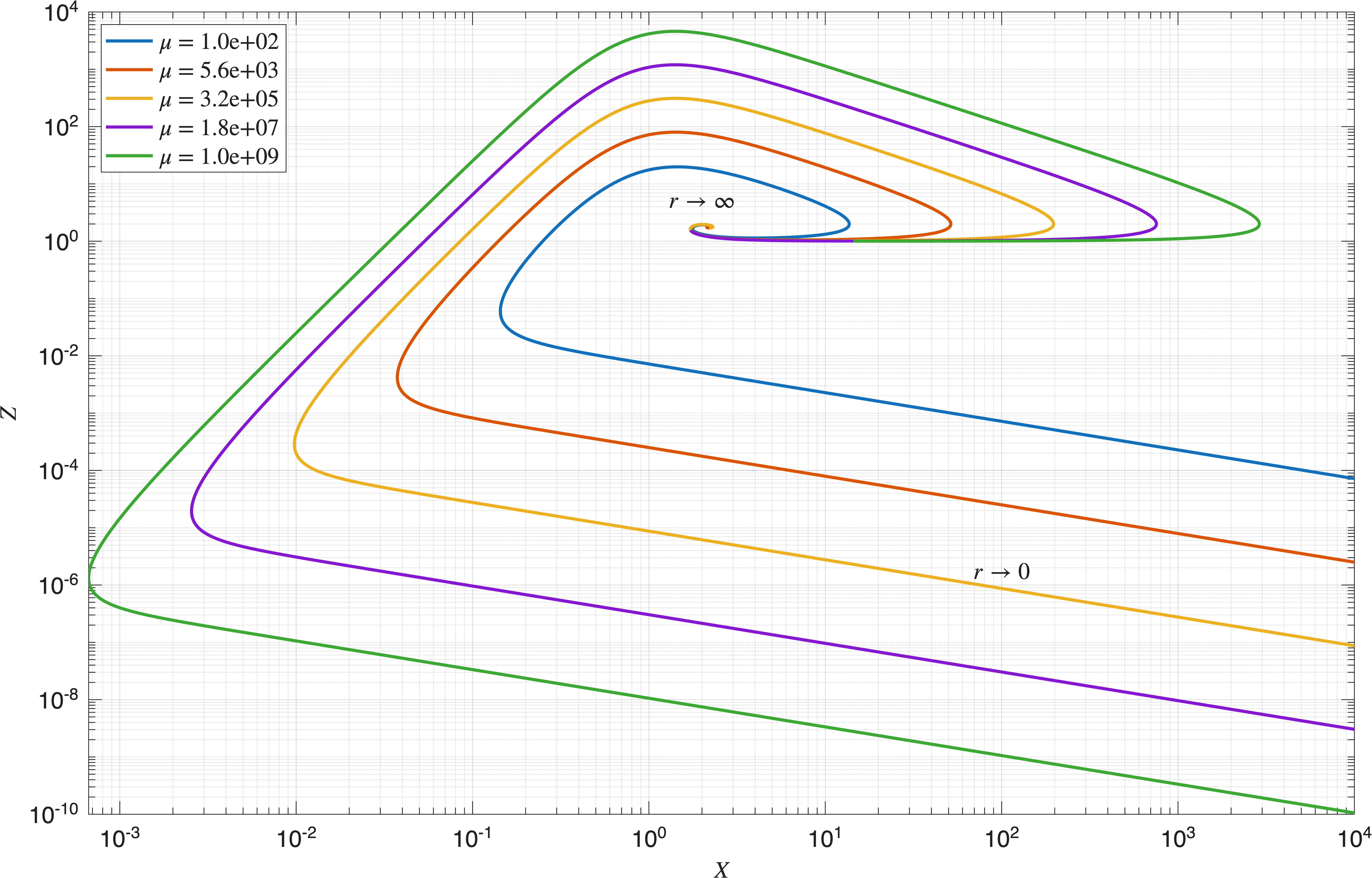}  % Adjust width
    \caption{Singular solutions from Fig.~\ref{fig:Fig1b}   in terms of variables $Z$ and $X$.}
    \label{fig:Fig2b}
\end{figure}

Figures~\ref{fig:Fig2a} and \ref{fig:Fig2b} display in the variables $Z$ and $X$ the solutions of Figs.~\ref{fig:Fig1a}
and~\ref{fig:Fig1b}, respectively.
The transition from the negative-mass
interior to the \splateau{} is
particularly evident in these variables. At large
radii, all solutions approach the same self-similar
asymptotic values, $Z\to 7/4$ and
$X\to\sqrt{14/3}$.

We now examine how the characteristic firehole
radius $R$ and mass $M$ depend on the central
negative mass parameter $\mu$.  With $A$ given by Eq.~(\ref{eq:A_q}), the characteristic scale $R_*^5=5\mu A^2$ becomes
\begin{equation}
R_*=\mu^{(1+2q)/5}.
\end{equation}
To quantify this behavior, we introduce the
dimensionless variables
\begin{equation}
\hat r=\frac{r}{R_*},
\qquad
\hat m=\frac{m}{R_*}.
\end{equation}
The firehole radius and mass can then be written as
\begin{equation}
R=\hat R R_*,
\qquad
M=\hat M R_*,
\end{equation}
where $\hat R=R/R_*$ and $\hat M=M/R_*$.
Numerically, these dimensionless quantities vary
only weakly with $\mu$ and approach constant
values over the large-$\mu$ range
examined. Denoting these limiting values by
$\hat R_\infty$ and $\hat M_\infty$, we obtain
\begin{equation}
\begin{aligned}
R &\simeq \hat R_\infty\,
\mu^{(1+2q)/5},\\
M &\simeq \hat M_\infty\,
\mu^{(1+2q)/5}.
\end{aligned}
\label{eq:RM_scaling}
\end{equation}
Thus, the predicted power-law exponent for these
families is
\begin{equation}
p_*(q)=\frac{1+2q}{5}.\label{eq:power-law}
\end{equation}
The scaling of $R_*$ follows directly from the
inner asymptotic expansion, whereas the
approximately constant ratios $R/R_*$ and $M/R_*$
are established numerically.
As shown in Sec.~\ref{sec:zero_family}, the limiting
values do not depend on $q$; numerically,
$\hat R_\infty\simeq 1.2457$ and
$\hat M_\infty\simeq 0.6229$.

The numerical results support these relations.
The power-law fits listed in Table~\ref{tab:tovf_scaling}
give, for $q=0$,
$R\simeq 1.311\,\mu^{0.196}$ and
$M\simeq 0.614\,\mu^{0.201}$, with fitted exponents
close to the predicted value $p_*(0)=1/5$; the same
agreement holds for the other values of $q$ in the table.
The compactness of the firehole approaches the
Schwarzschild value: $M/R\simeq 0.482$ at $\mu=10^2$
for $q=0$, and $M/R\to 1/2$ as $\mu\to\infty$ for $q<2$, so that
the minimum of $1-2m/r$ at $R$ tends to zero, although
the solutions contain no event horizon at any finite $\mu$.

The choice $q=1/3$ has a particularly simple
interpretation. Near the origin, the lapse
satisfies $e^{2\nu}\simeq A/r$. Expressing this
relation in terms of $\hat r=r/R_*$ gives
\begin{equation}
e^{2\nu}\simeq
\frac{1}{\sqrt{5}\,\hat r}
\mu^{(3q-1)/5}.
\label{eq:inner_lapse_scaling}
\end{equation}
Hence, $q=1/3$ is the unique choice for which
the leading inner lapse profile, expressed in
terms of $\hat r$, becomes independent of $\mu$.
This provides an interpretation of the
$\mu^{1/3}$ scaling associated with the
Zurek--Page family: it corresponds to comparing
solutions with a common dimensionless inner
lapse normalization.

More generally, the TOV equations do not
determine a unique relationship between the
central negative mass $\mu$ and the
characteristic firehole mass $M$. Such a
relationship arises only after specifying
how the remaining integration constant $A$
is chosen along a family of solutions.
Up to the scaling symmetry discussed in
Sec.~\ref{sec:zero_family}, the singular solutions
form a one-parameter family labeled by
$\beta=\mu/R_*$, and the choice of $q$ only fixes
how $\beta$ grows with $\mu$.

\section{Radial Stability Analysis}
\label{sec:4}

We study radial oscillations using the formulation of
Kokkotas and Ruoff~\cite{kostas_radial}, based on
Chandrasekhar's original treatment~\cite{Chandrasekhar1964ApJ,
Chandrasekhar1964PRL}.

Starting with the time-dependent radial displacement $\delta r(t,r)$ of a fluid element located at position $r$, one introduces a harmonic time dependence of the form
\begin{equation}
    \delta r(t,r) = \xi(r)\,e^{i\,\omega\,t} = \zeta(r)\,e^\nu\,e^{i\,\omega\,t}/r^2\,.\label{eq:dr}
\end{equation}
The master equation for $\zeta$ reads~\cite{kostas_radial}
\begin{equation}
    \left(P\zeta'\right)' + (Q + \omega^2 W)\zeta = 0\,,
    \label{eq:RadOsc}
\end{equation}
which, for convenience, one rewrites as
\begin{eqnarray}
    \zeta' &=& \frac{\eta}{P}, \label{eq:RO1} \\
    \eta' &=& -(Q+\omega^2 W)\zeta\,.\label{eq:RO2}
\end{eqnarray}
The coefficients are
\begin{eqnarray}
    r^2 W &=& (\rho+p)\, e^{3\lambda + \nu}, \label{eq:W}\\
    r^2 P &=& \gamma\,p\, e^{\lambda + 3\nu},\label{eq:P} \\
    r^2 Q &=& (\rho+p)\,e^{\lambda+3\nu}
    \left[\nu'^2+\frac{4}{r}\nu'-8\pi p\,e^{2\lambda}\right],\label{eq:Q}
\end{eqnarray}
where
\begin{equation}
    \gamma = \frac{\rho+p}{p}\frac{dp}{d\rho}\,.
\end{equation}
With our \eos{} $p = \rho/3$ and our normalization $4\,\pi\rho\,e^{4\,\nu} = 1$, these coefficients can be rewritten as
\begin{eqnarray}
    r^2 W &=&  e^{3(\lambda - \nu)}, \label{eq:Wc}\\
    r^2 P &=&  \frac{1}{3}\, e^{\lambda - \nu},\label{eq:Pc} \\
    r^2 Q &=&  e^{\lambda -\nu} \,
    \left[ \left(\nu'\right)^2 + \frac{4}{r}\nu' - \frac{2}{3} \, e^{2\lambda-4\,\nu}  \right]\,,\label{eq:Qc}
\end{eqnarray}
where we have absorbed a factor of $3\,\pi$ in the coefficients.

We solve Eqs.~(\ref{eq:RO1})--(\ref{eq:RO2}) on $[R_0,R_{\rm out}]$ as an eigenvalue
problem for $\omega^2$. The radiation fluid has $p(R_{\rm out})>0$
at every finite $R_{\rm out}$, so $R_{\rm out}$ is a chosen boundary
of the perturbation problem rather than the surface of a star
matched directly to a vacuum.
Truncating the equilibrium at $R_{\rm out}$ therefore presupposes
an exterior agent that supplies the pressure $p(R_{\rm out})$. The
two outer conditions introduced below correspond to two idealized
agents: a constant-pressure bath, which leaves the Lagrangian
pressure of the boundary fluid element unchanged, and a rigid wall
held at fixed areal radius $R_{\rm out}$, which prevents the boundary
element from moving.

The initial conditions at $R_0$ are derived as follows. We found in Sec.~\ref{sec:equilibria} that as $r \rightarrow 0$, $m \rightarrow -\mu$  and
\begin{eqnarray}
e^{2\,\nu} &\simeq& \frac{A}{r}\\
e^{2\,\lambda} &\simeq& \left(1+\frac{2\,\mu}{r}\right)^{-1} \simeq \frac{r}{2\,\mu}\,.
\end{eqnarray}
Notice that with our choice $A = \mu^q/\sqrt{5}$
\begin{equation}
    e^{\lambda-\nu} = E\,r\,,
\end{equation}
where
\begin{equation}
    E^{-2} \equiv 2\,\mu\,A = \sqrt{\frac{4}{5}}\,\mu^{q+1}\,.
\end{equation}
With these approximations, the coefficients $W$, $P$, and $Q$ read
\begin{eqnarray}
     W &\simeq& r\,E^3 \\
     P &\simeq& \frac{E}{3\,r} \\
     Q &\simeq& -\frac{7\,E}{4\,r^3}\,.
\end{eqnarray}
Substitution of these approximations into Eq.~(\ref{eq:RadOsc}) yields
\begin{equation}
   \left(\frac{\zeta'}{r}\right)' - \frac{21}{4}\,\frac{\zeta}{r^3} =0 \,,
    \label{eq:RadOsc2}
\end{equation}
where we have neglected the term with the $W$ coefficient because $W  = {\cal O}(r)$ compared with $Q = {\cal O}(r^{-3})$. Hence,
\begin{equation}
    \zeta = c_1\, r^{7/2} + c_2\, r^{-3/2}\,.
\end{equation}

Recall that $\delta r=e^\nu\zeta/r^2$. Since $e^\nu\sim r^{-1/2}$, the regular branch $\zeta\sim r^{7/2}$ gives $\delta r\sim r$, whereas the second branch $\zeta\sim r^{-3/2}$ gives $\delta r\sim r^{-4}$. The latter, therefore, corresponds to a divergent physical displacement and must be discarded. Hence, the physically relevant solution is $\zeta=c_1r^{7/2}$.
The same selection follows from the natural
Sturm--Liouville norm. Since $W\sim r$ near
the origin, the regular branch satisfies
$W\zeta^2\sim r^8$, whereas the singular
branch gives $W\zeta^2\sim r^{-2}$.
Only the former is integrable at $r=0$.
The singular endpoint is therefore of
limit-point type, so no additional independent
boundary condition is required there beyond
selecting the admissible solution.

Since the radial perturbation equation is defined up to a rescaling constant on $\zeta$, we will also set $c_1=1$. Finally from Eq.~(\ref{eq:RO1}),
\begin{equation}
    \eta = \zeta'\, P = \frac{7\,E}{6}\,r^{3/2}\,.
\end{equation}
The initial conditions at $R_0$ will then be
\begin{eqnarray}
\zeta &=& R_0^{7/2}\label{eq:zeta0}\\
 \eta &=& \zeta'\, P = \frac{7\,E}{6}\,R_0^{3/2}.\label{eq:eta0}
\end{eqnarray}

For the outer boundary, we will consider two situations: a \ac{tovf} with an open boundary, referred to as \ac{tovfo}, and a \ac{tovf} with a rigid wall, referred to as \ac{tovfb}.
For the \ac{tovfo}, we impose
$\eta(R_{\rm out})=0$, equivalently $\Delta p(R_{\rm out})=0$, and for the \ac{tovfb}, we require that $\zeta(R_{\rm out}) = 0$.

At each outer radius $R_{\rm out}$, we follow the nodeless, fundamental radial
mode. We denote by $R_c$ the critical radius at which this mode
has $\omega^2=0$; its eigenvalue is positive below and negative
above this crossing in the range examined. For the boxed problem
this crossing is unique for every $R_{\rm out}$, because
$\omega^2_{0,\rm B}$ decreases monotonically with $R_{\rm out}$
(Sec.~\ref{sec:bounds}).

To determine the fundamental eigenvalue, we employ the
Pr\"ufer phase method, which provides a convenient way
to identify the radial eigenmodes by their number of
nodes. This is particularly important because a direct
shooting method that enforces the outer boundary
condition does not, by itself, distinguish the fundamental
mode from higher overtones.

We introduce the Pr\"ufer variables $b(r)$ and
$\theta(r)$ through
\begin{equation}
    \zeta = b\sin\theta,\qquad
    \eta = \kappa b\cos\theta,
\end{equation}
where $\kappa>0$ is an arbitrary constant that can be
chosen to improve numerical conditioning. Substituting
these expressions into Eqs.~(\ref{eq:RO1}) and (\ref{eq:RO2}), we obtain
\begin{equation}
    \theta' =
    \frac{\kappa}{P}\cos^2\theta
    +\frac{Q+\omega^2 W}{\kappa}\sin^2\theta.
    \label{eq:prufer}
\end{equation}
The phase equation is independent of the amplitude $b$,
so the eigenvalues can be determined without integrating
the potentially rapidly varying displacement amplitude.

The regular inner solution, Eqs.~(\ref{eq:zeta0}) and (\ref{eq:eta0}),
determines the initial phase,
\begin{equation}
    \theta(R_0)=
    \arctan\left(\frac{6\kappa R_0^2}{7E}\right).
\end{equation}
The two outer boundary conditions become phase
conditions. For TOVFO, $\eta(R_{\rm out})=0$ requires
\begin{equation}
    \theta(R_{\rm out})=\frac{\pi}{2}+n\pi,
    \label{eq:prufer_open}
\end{equation}
whereas for TOVFB, $\zeta(R_{\rm out})=0$ requires
\begin{equation}
    \theta(R_{\rm out})=n\pi.
    \label{eq:prufer_box}
\end{equation}

The integer $n$ labels the phase branches and is
related to the number of interior nodes of $\zeta$.
In particular, the fundamental mode has no interior
nodes and, with our choice of initial phase, satisfies
$\theta(R_{\rm out})=\pi/2$ for TOVFO and
$\theta(R_{\rm out})=\pi$ for TOVFB.

For each background solution, we identify the
critical radius $R_c$ by integrating the
zero-frequency Pr\"ufer equation and locating
the first radius at which the corresponding
fundamental-mode phase condition is satisfied.
For TOVFO, this occurs at $\theta=\pi/2$,
whereas for TOVFB it occurs at $\theta=\pi$.
These radii define the marginal-stability
thresholds, at which $\omega_0^2=0$,
separating the stable and unstable branches
of the fundamental radial mode.

\begin{table*}[t]
\centering
\caption{Power-law fits $X=C_X\mu^{p_X}$ for the firehole radius $R$, plateau mass $M$, and critical radius $R_c$ for open (O) and boxed (B) boundary conditions. Fits use $10\leq\mu\leq10^7$, sampled at logarithmically spaced values. The predicted exponent is $p_*(q)=(1+2q)/5$.}
\label{tab:tovf_scaling}
\begin{ruledtabular}
\begin{tabular}{c c cc cc cc cc}
$q$ & $p_*$ & $C_R$ & $p_R$ & $C_M$ & $p_M$ & $C_c^{\rm O}$ & $p_c^{\rm O}$ & $C_c^{\rm B}$ & $p_c^{\rm B}$ \\
\hline
0 & 0.200 & 1.3110 & 0.1959 & 0.6137 & 0.2011 & 1.1433 & 0.1982 & 1.3124 & 0.1958 \\
0.25 & 0.300 & 1.3306 & 0.2948 & 0.6138 & 0.3010 & 1.1528 & 0.2975 & 1.3324 & 0.2946 \\
0.5 & 0.400 & 1.3574 & 0.3933 & 0.6148 & 0.4007 & 1.1665 & 0.3966 & 1.3601 & 0.3931 \\
0.75 & 0.500 & 1.3947 & 0.4913 & 0.6173 & 0.5002 & 1.1869 & 0.4953 & 1.3992 & 0.4911 \\
1 & 0.600 & 1.4491 & 0.5888 & 0.6229 & 0.5992 & 1.2179 & 0.5934 & 1.4557 & 0.5884 \\
\end{tabular}
\end{ruledtabular}
\end{table*}

Table~\ref{tab:tovf_scaling} shows coefficients of power-law fits $X=C_X\mu^{p_X}$ for the firehole radius $R$, plateau mass $M$, and critical radius $R_c$ for open (O) and boxed (B) boundary conditions. Fits use $10\leq\mu\leq10^7$. The predicted exponent from Eq.~(\ref{eq:power-law}) is $p_*(q)=(1+2q)/5$.

The numerical results reveal that the critical radius
shares approximately the same scaling with the central
mass parameter as the equilibrium firehole radius and
mass. As shown in Table~\ref{tab:tovf_scaling}, the fitted exponents for
$R_c$, for both boundary conditions, are in close
agreement with the prediction
$p_*(q)=(1+2q)/5$ of Eq.~(\ref{eq:power-law}). Thus, the onset
of radial instability is governed by a characteristic
length scale that follows the equilibrium scaling.
The agreement extends over the range
$10\leq\mu\leq 10^7$ and for all the values of $q$
examined. The small, $q$-dependent departures of the
fitted exponents from $p_*$ reflect the slow approach
of $R/R_*$ to its limiting value, which, as shown in
Sec.~\ref{sec:zero_family}, is controlled by
$\beta=\mu^{(4-2q)/5}$ and is therefore slower for
larger $q$.

A more interesting result emerges from the ratio of the
critical radius to the firehole radius. For TOVFO, the
critical radius is systematically smaller than the
firehole radius, whereas for TOVFB the two radii nearly
coincide. Over the range examined,
$0.83\lesssim R_c^{\rm O}/R\lesssim 0.898$, with the lower
end attained at the smallest $\beta$ ($q=1$, $\mu=10$), while
$R_c^{\rm B}/R$ differs from unity by less than
$10^{-2}$ (Fig.~\ref{fig:Rc_vs_beta}).
These results indicate that the location of the
marginal-stability boundary depends sensitively
on the outer boundary condition. In particular,
confining the radial displacement at the outer
boundary allows the configuration to remain stable
up to a radius very close to the characteristic
firehole radius $R$.

\subsection{Fundamental zero-mode eigenfunctions}
\label{sec:zero_modes}

\begin{figure}[htbp]
    \centering
    \includegraphics[width=\linewidth]{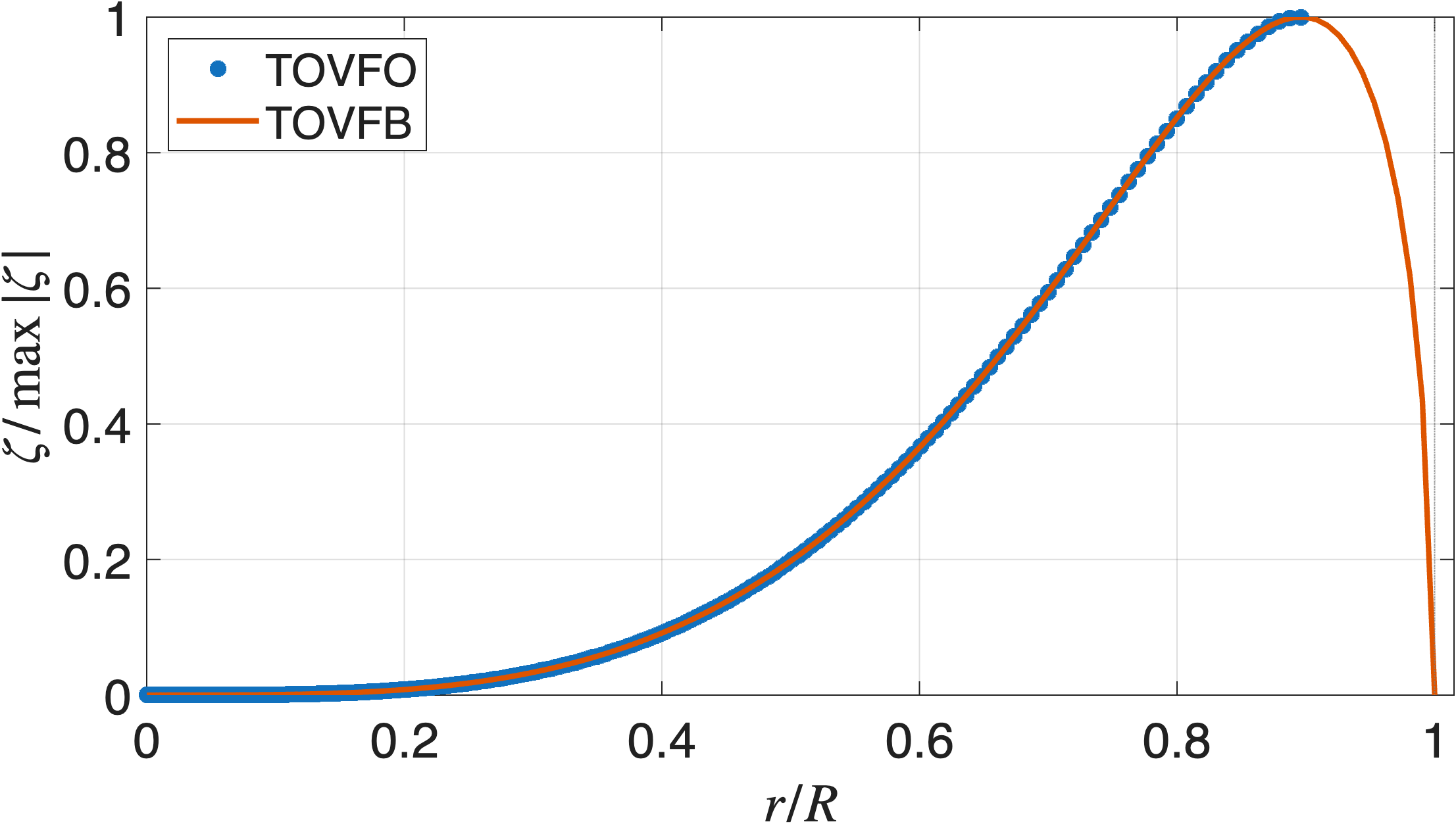}  % Adjust width
    \caption{Normalized radial displacement
$\zeta/\max|\zeta|$ for the fundamental
zero modes of TOVFO and TOVFB, computed
on the same equilibrium background and
terminated at their respective critical
radii. The dotted vertical line marks
$r=R$.}
    \label{fig:zero_zeta}  % Label for referencing
\end{figure}

To examine the distinction between the two boundary
conditions, we compare the zero-frequency solutions
evaluated at their respective critical radii,
$R_c^{\rm O}$ and $R_c^{\rm B}$, for the same TOVF
background. Figure~\ref{fig:zero_zeta} shows the displacement variable
$\zeta$, normalized by its maximum absolute value.
The two zero modes are obtained from the same regular
solution of the zero-frequency pulsation equations,
truncated at their respective critical radii.
Their agreement over the common radial interval is
therefore expected; any small differences in the
plotted curves arise from the independent amplitude
normalizations and numerical accuracy. For TOVFO,
the solution terminates at $\eta(R_c^{\rm O})=0$,
whereas for TOVFB it terminates at
$\zeta(R_c^{\rm B})=0$.

\begin{figure}[htbp]
    \centering
    \includegraphics[width=\linewidth]{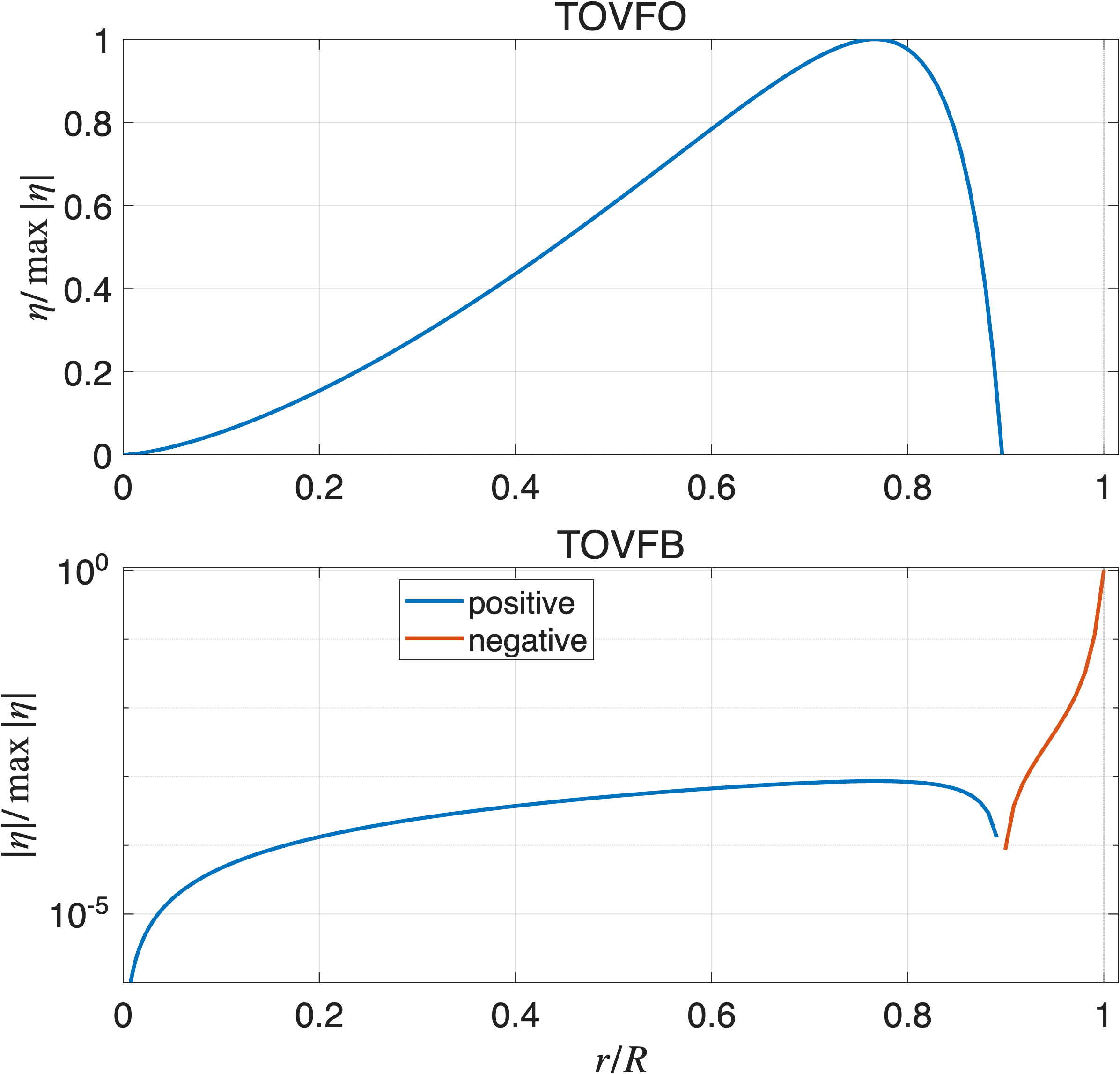}  % Adjust width
    \caption{Momentum-like variable $\eta=P\zeta'$ for the fundamental zero
 modes: TOVFO (upper panel) and TOVFB (lower panel). Each curve is
 normalized by its own $\max|\eta|$ to display its radial structure; the relative magnitudes between panels are therefore not shown.
 The dotted vertical line marks $r=R$.}
    \label{fig:zero_eta}  % Label for referencing
\end{figure}

The momentum-like variable $\eta=P\zeta'$ is shown separately for each
boundary condition in Fig.~\ref{fig:zero_eta}. Independent normalization
by $\max|\eta|$ is used to resolve the shapes. In the open problem,
$\eta$ reaches zero at $R_c^{\rm O}$. In the boxed problem, the
vanishing of $\zeta$ at $R_c^{\rm B}$ is instead accompanied by a
nonzero $\eta$. Its rapid variation near the endpoint should not be
interpreted as a discontinuity or a distinct inner eigenfunction.

\begin{figure}[htbp]
    \centering
    \includegraphics[width=\linewidth]{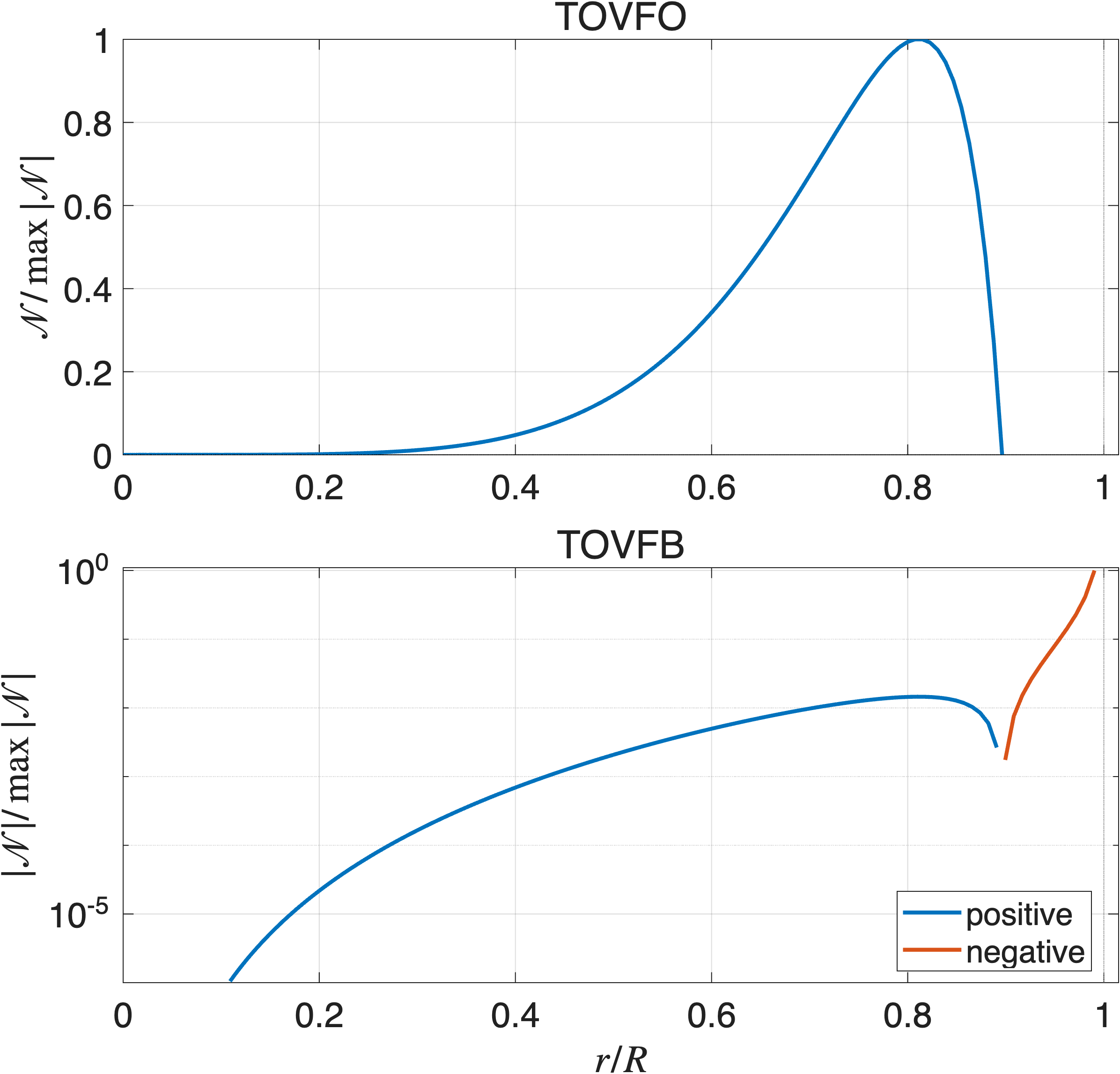}  % Adjust width
    \caption{Cumulative zero-mode variational numerator $\mathcal N(r)$ for TOVFO (upper panel) and TOVFB (lower panel), independently normalized
by $\max|\mathcal N|$. The identity in
 Eq.~\eqref{eq:zero_N_identity} provides a direct consistency check.
The dotted vertical line marks $r=R$.}
    \label{fig:zero_N}  % Label for referencing
\end{figure}

At zero frequency, Eqs.~(\ref{eq:RO1})--(\ref{eq:RO2}) imply the identity
\begin{equation}
 (\zeta\eta)'=\frac{\eta^2}{P}-Q\zeta^2.
\end{equation}
Consequently, the cumulative variational numerator measured from the
inner integration radius is
\begin{eqnarray}
 \mathcal N(r)&=&\int_{R_0}^{r}\!ds\,
 \left[\frac{\eta^2}{P}-Q\zeta^2\right]\nonumber\\
 &=&\zeta(r)\eta(r)-\zeta(R_0)\eta(R_0).
 \label{eq:zero_N_identity}
\end{eqnarray}
The endpoint term vanishes as $R_0\to0$ for the regular inner branch.
The two-panel comparison in Fig.~\ref{fig:zero_N} illustrates how
$\mathcal N(r)$ accumulates in the respective zero modes. Its value
vanishes at the critical radius for either condition, because the
boundary product $\zeta\eta$ vanishes. The sign of this cumulative
quantity at an intermediate radius must not be construed as the
stability of an independently truncated configuration unless its boundary condition is also satisfied there.

\subsection{Variational interpretation of the stability
thresholds}
\label{sec:variational}

The different critical radii obtained for TOVFO and
TOVFB can be understood in terms of the variational
properties of the radial pulsation operator. Multiplying
Eq.~(\ref{eq:RadOsc}) by $\zeta$ and integrating by parts yields
\begin{equation}
\omega^2 =
\frac{
\displaystyle
\int_0^{R_{\rm out}}
\left[P(\zeta')^2-Q\zeta^2\right]dr
}{
\displaystyle
\int_0^{R_{\rm out}}W\zeta^2\,dr
},
\label{eq:rayleigh}
\end{equation}
where the boundary term vanishes for both problems, and the
lower limit is understood as $R_0\to0$ along the regular branch, as in
Eq.~\eqref{eq:zero_N_identity}.
At the inner endpoint, the regular solution satisfies
$\zeta\eta\sim r^5\to0$. At the outer endpoint,
the boundary term vanishes because $\eta(R_{\rm out})=0$
for TOVFO and $\zeta(R_{\rm out})=0$ for TOVFB.
Since $P>0$ and $W>0$, Eq.~\eqref{eq:rayleigh}
provides the Rayleigh quotient associated with the
radial pulsation problem.

The fundamental eigenvalue is characterized by the
minimum of this quotient over the admissible radial
perturbations. The two boundary conditions differ
in their admissible function spaces. For TOVFO,
$\eta(R_{\rm out})=0$ is the natural boundary condition
associated with the variational problem, whereas
TOVFB imposes the additional restriction
$\zeta(R_{\rm out})=0$. Consequently, the admissible
variations for TOVFB form a subset of those for
TOVFO. For the same equilibrium background and
outer radius $R_{\rm out}$, the variational principle gives
\begin{equation}
\omega_{0,\mathrm{O}}^2(R_{\rm out})
\leq
\omega_{0,\mathrm{B}}^2(R_{\rm out}).
\label{eq:eigenvalue_order}
\end{equation}
Here the subscripts O and B denote the open and
boxed problems, respectively. This ordering assumes
the same regular inner boundary condition and the
corresponding self-adjoint realization of the
pulsation operator.

Equation~\eqref{eq:eigenvalue_order} establishes
that the rigid outer boundary cannot lower the
fundamental squared frequency. Thus, confining the
fluid displacement has a stabilizing effect. In
particular, if the fundamental eigenvalue of TOVFO
is positive at a given outer radius, the corresponding
boxed configuration must also be stable against radial
perturbations.

The ordering also has a direct interpretation in
terms of the zero-frequency solution. As discussed
in Sec.~\ref{sec:zero_modes}, the regular zero mode is common to
both boundary-value problems. Starting at the
origin, $\zeta$ and $\eta$ are positive. The first
zero of $\eta$, corresponding to an extremum of
$\zeta$, determines $R_c^{\rm O}$. Continuing the
same solution outward, $\eta$ becomes negative and
$\zeta$ decreases until its first zero at
$R_c^{\rm B}$. Therefore,
\begin{equation}
R_c^{\rm O}<R_c^{\rm B},
\label{eq:critical_order}
\end{equation}
in agreement with our numerical results.

The ordering in Eq.~(\ref{eq:eigenvalue_order}) holds at fixed
$R_{\rm out}$ for the same equilibrium background.
It does not require the two eigenfunctions
to have the same radial profile. At zero
frequency, however, their critical solutions
are different truncations of the same
regular solution, which establishes the
ordering of the critical radii.

The cumulative variational numerator introduced in
Eq.~(\ref{eq:zero_N_identity}) provides another way to understand this
behavior. In the limit $R_0\to0$, the zero-frequency
identity gives
\begin{equation}
\mathcal{N}(r)=\zeta(r)\eta(r).
\end{equation}
For the fundamental zero mode, $\zeta$ remains positive
throughout the interior. Hence, $\mathcal{N}$ is
positive up to $R_c^{\rm O}$, vanishes at the first
zero of $\eta$, becomes negative between the two
critical radii, and returns to zero when $\zeta$
vanishes at $R_c^{\rm B}$. This behavior is illustrated
in Fig.~\ref{fig:zero_N}.

The negative cumulative numerator in the interval
$R_c^{\rm O}<r<R_c^{\rm B}$ does not imply that the
boxed configuration is unstable. Indeed, at
$R_{\rm out}=R_c^{\rm B}$, the same zero-mode solution
satisfies the rigid-wall boundary condition and has
$\omega_0^2=0$. Stability is determined by the
eigenvalue of the complete boundary-value problem,
rather than by the sign of the cumulative integral
at an intermediate radius.

\subsection{Zero modes as neighboring equilibria}
\label{sec:zero_family}

The near-coincidence $R_c^{\rm B}\simeq R$ has a simple origin: the regular
zero-frequency solution is the displacement that connects neighboring members of
the equilibrium family of Sec.~\ref{sec:equilibria}.

Consider a one-parameter family of TOVF equilibria, each a solution of the TOV
Eqs.~\eqref{eq:tov1a} and \eqref{eq:tov2a} with
the same constant $K$, and let $\delta$ denote the derivative with respect to the
family parameter at fixed $r$. Because every member solves the TOV equations, the
variations $\delta m$, $\delta\rho$, $\delta p$, $\delta\nu$ satisfy their
linearization. To compare with the pulsation problem, we describe the passage
from one member to its neighbor as a radial displacement of the fluid,
$\delta r=\xi(r)$, which is the displacement of Eq.~\eqref{eq:dr} at $\omega=0$.
We define $\xi$ through the change in the enclosed mass at fixed $r$,
\begin{equation}
  \delta m=-4\pi r^{2}(\rho+p)\,\xi ,
  \label{eq:dm_xi}
\end{equation}
or equivalently $\delta\lambda=-(\nu'+\lambda')\,\xi$. In the derivation of
Eq.~\eqref{eq:RadOsc}, this relation follows from the $tr$ component of the
perturbed Einstein equations, $\partial_t\delta\lambda=-(\nu'+\lambda')\,\partial_t\xi$,
integrated in time from the unperturbed state. Combined with
$\delta m=r\,e^{-2\lambda}\delta\lambda$, which follows from
$e^{-2\lambda}=1-2m/r$, and the background identity
$\nu'+\lambda'=4\pi r(\rho+p)\,e^{2\lambda}$, it gives Eq.~\eqref{eq:dm_xi},
which therefore holds for any equation of state: the change in $m$ at fixed
$r$ is the mass-energy carried across $r$ by the displaced fluid. Between
static equilibria there is no time evolution, and Eq.~\eqref{eq:dm_xi} serves
as the definition of $\xi$, chosen to coincide with the $\omega\to0$ limit of
the displacement in Eq.~\eqref{eq:RadOsc}.

Varying Eq.~\eqref{eq:TOV1} gives $\delta m'=4\pi r^{2}\delta\rho$. Together with
Eq.~\eqref{eq:dm_xi} and $p'=-(\rho+p)\nu'$ from Eqs.~\eqref{eq:TOV2}
and~\eqref{eq:TOV3}, this yields the Lagrangian density perturbation
\begin{eqnarray}
  \Delta\rho&\equiv&\delta\rho+\rho'\xi
  =-(\rho+p)\Big[\frac{1}{r^{2}}\big(r^{2}\xi\big)'-\nu'\xi\Big]\nonumber\\
  &=&-(\rho+p)\,\frac{e^{\nu}}{r^{2}}\,\zeta',
  \label{eq:Delta_rho}
\end{eqnarray}
where the last step uses Eq.~\eqref{eq:dr}. The variations of
Eqs.~\eqref{eq:TOV2} and~\eqref{eq:TOV3} hold automatically, since each member
of the family is an equilibrium. Because every member obeys $p=\rho/3$, the
Lagrangian perturbations satisfy $\Delta p=\Delta\rho/3=\gamma p\,\Delta\rho/(\rho+p)$
with the same $\gamma=4/3$ that enters $P$. This is the adiabatic relation
assumed in the pulsation equation. The correspondence between neighboring
equilibria and zero modes therefore requires the adiabatic index of the
perturbations to equal $(\rho+p)\,p^{-1}dp/d\rho$ evaluated along the
equilibrium family, i.e., $\gamma=\Gamma$. For thermal radiation this holds
exactly, since adiabatic compression of a photon gas preserves $p=\rho/3$.
With this relation, $\Delta p=-\gamma\,p\,e^{\nu}\zeta'/r^{2}$, which is the
relation between $\Delta p$ and $\zeta$ used in deriving Eq.~\eqref{eq:RadOsc},
and the linearized hydrostatic equilibrium, Eq.~\eqref{eq:TOV2}, reduces to
Eq.~\eqref{eq:RadOsc} at $\omega=0$. Finally, using $\rho+p=e^{-4\nu}/(3\pi)$, which
follows from $4\pi\rho\,e^{4\nu}=1$ and $p=\rho/3$, Eqs.~\eqref{eq:dr}
and~\eqref{eq:dm_xi} give
\begin{equation}
  \zeta_\delta\equiv r^{2}e^{-\nu}\xi=-\frac34\,e^{3\nu}\,\delta m ,
  \label{eq:zeta_delta}
\end{equation}
which solves Eq.~\eqref{eq:RadOsc} with $\omega=0$. For a general
$\Gamma$-law fluid, Eq.~\eqref{eq:TOV3} integrates to
$4\pi\rho\,e^{\Gamma\nu/(\Gamma-1)}=K$, and the same steps give
$\zeta_\delta=-(\Gamma K)^{-1}e^{\nu/(\Gamma-1)}\,\delta m$, which reduces to
Eq.~\eqref{eq:zeta_delta} for $\Gamma=4/3$ and $K=1$. This is the
general-relativistic form of the classic correspondence between zero-frequency
modes and neighboring equilibria~\cite{Harrison1965}. In the same normalization
(with the factor $3\pi$ absorbed into $P$ as above), the momentum-like variable is
the Lagrangian pressure perturbation,
$\eta_\delta=P\zeta_\delta'=-3\pi e^{\lambda+2\nu}\Delta p$, with
\begin{equation}
  \Delta p=\frac{1}{12\pi r^{2}}\Big(3\nu'\,\delta m-4r^{2}e^{-4\nu}\delta\nu\Big).
  \label{eq:Dp}
\end{equation}

The inner expansion, Eq.~\eqref{eq:mcore}, identifies which family derivative is
regular. Varying the inner lapse normalization at fixed central mass gives
$\partial_A m|_\mu=-2r^{5}/(5A^{3})$ near the origin, so
$\zeta_\delta\propto e^{3\nu}r^{5}\propto r^{7/2}$, i.e. $\xi\propto r$: the
regular branch. Varying the central mass at fixed $A$ gives
$\partial_\mu m|_A=-1$, so $\zeta_\delta\propto r^{-3/2}$, i.e. $\xi\propto r^{-4}$:
the discarded singular branch. The singular solution therefore corresponds to a
change of the central point mass, which a regular fluid perturbation cannot
produce. Since the regular zero-frequency solution is unique up to normalization,
\begin{equation}
  \zeta_0(r)\propto e^{3\nu}\,\frac{\partial m}{\partial A}\Big|_{\mu,K}.
  \label{eq:zeta0_family}
\end{equation}
We verified Eq.~\eqref{eq:zeta0_family} numerically: the directly integrated
zero-frequency solution and the finite-difference family derivative agree to
better than $10^{-7}$ in relative amplitude over the full radial range, for all
backgrounds examined.

Equation~\eqref{eq:zeta0_family} gives both thresholds a turning-point interpretation.
For TOVFB, $\zeta(R_{\rm out})=0$ (equivalently $\xi(R_{\rm out})=0$) is equivalent to
\begin{equation}
  \frac{\partial m(R_{\rm out})}{\partial A}\Big|_{\mu,K}=0 ,
  \label{eq:turning}
\end{equation}
so $R_c^{\rm B}$ is the radius at which the mass enclosed by the wall is
stationary along the family of equilibria with fixed central mass and fixed $K$.
This is the analog, for a confined configuration, of the familiar criterion that
the fundamental radial mode of a star changes stability at an extremum of the mass
along the equilibrium sequence~\cite{Harrison1965,BardeenThorneMeltzer1966,Sorkin1982}. For TOVFO, $\eta(R_{\rm out})=0$ is equivalent to
$\Delta p(R_{\rm out})=0$ in Eq.~\eqref{eq:Dp}: the Lagrangian pressure of the boundary
fluid element is stationary along the same family.

\begin{figure}[htbp]
  \centering
  \includegraphics[width=\linewidth]{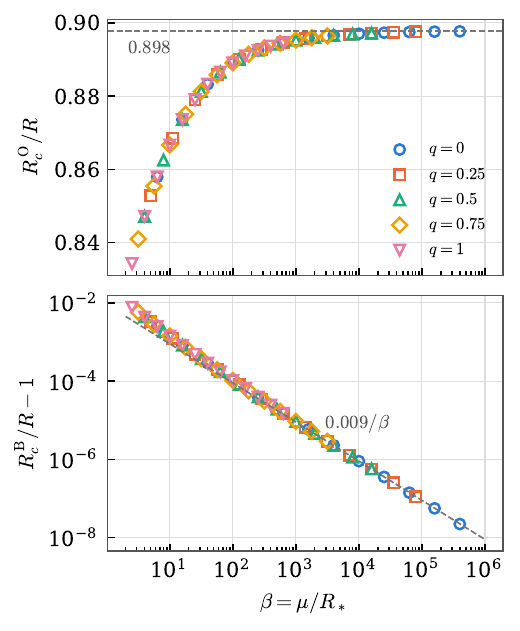}
  \caption{Critical radii for $0\le q\le1$ and
  $10\le\mu\le10^7$, plotted against the scale-invariant parameter
  $\beta=\mu/R_*$ of Eq.~(\ref{eq:beta}). Top: open boundary,
  $R_c^{\rm O}/R$; the dashed line marks the large-$\beta$ limit
  $0.898$. Bottom: boxed boundary, $R_c^{\rm B}/R-1$; the dashed line
  is $0.009/\beta$. Results for different $q$ fall on a single curve.}
  \label{fig:Rc_vs_beta}
\end{figure}

\paragraph*{Scaling symmetry and the origin of $R_c^{\rm B}\simeq R$.}
At fixed $K$, the TOV equations in the form of Eqs.~\eqref{eq:tov1a}
and~\eqref{eq:tov2a} are invariant under
\begin{equation}
  r\to s\,r,\qquad m\to s\,m,\qquad y\to s\,y,
  \label{eq:scaling}
\end{equation}
which maps $\mu\to s\mu$, $A\to s^{2}A$, and $R_*\to sR_*$. Up to this rescaling,
the singular solutions form a one-parameter family labeled by the scale-invariant
ratio
\begin{equation}
  \beta\equiv\frac{\mu}{R_*}=\Big(\frac{\mu^{4}}{5A^{2}}\Big)^{1/5}=\mu^{(4-2q)/5},
  \label{eq:beta}
\end{equation}
and the mass function takes the form
\begin{equation}
  m(r;\mu,A)=R_*\,F(x;\beta),\qquad x=r/R_* .
  \label{eq:F}
\end{equation}
Every dimensionless ratio, including $R/R_*$, $M/R_*$, and $R_c/R$, depends on
$\mu$ and $q$ only through $\beta$; the exponent $q$ merely selects the path
$\beta(\mu)$. Using $\partial\ln R_*/\partial\ln A|_\mu=2/5$ and
$\partial\ln\beta/\partial\ln A|_\mu=-2/5$,
\begin{equation}
  A\,\frac{\partial m}{\partial A}\Big|_\mu
  =-\frac25\Big[r^{2}\Big(\frac{m}{r}\Big)'+R_*\,\beta\,\frac{\partial F}{\partial\beta}\Big].
  \label{eq:dmdA}
\end{equation}
The approach of $R/R_*$ and $M/R_*$ to constant values found in
Sec.~\ref{sec:equilibria} expresses that, outside the interior core, $F$ becomes
independent of $\beta$ as $\beta\to\infty$. In that limit, the zero of
Eq.~\eqref{eq:dmdA} is the zero of $(m/r)'$, which is the definition of the
firehole radius $R$. Hence
\begin{equation}
  \lim_{\beta\to\infty}\frac{R_c^{\rm B}}{R}=1 .
  \label{eq:limit}
\end{equation}
The relation $R_c^{\rm B}\simeq R$ is therefore not a numerical coincidence: the
rigid-wall zero mode vanishes where the equilibrium compactness $m/r$ is
extremal. Figure~\ref{fig:Rc_vs_beta} confirms both the collapse onto a single function of
$\beta$ and the approach to this limit, with $R_c^{\rm B}/R-1\simeq0.009/\beta$ for
large $\beta$. No comparable closed form follows for the open threshold;
numerically, $R_c^{\rm O}/R\to0.898$.

The $1/\beta$ approach of $R_c^{\rm B}/R$ is faster than the
convergence of the equilibrium scales themselves: numerically,
$R/R_*-\hat R_\infty$ decreases only roughly as $\beta^{-1/2}$. The
argument leading to Eq.~\eqref{eq:limit} establishes the limit but not
this rate. The faster convergence is consistent with the leading
large-$\beta$ correction to $F$ acting approximately as a rescaling of
$x$, $F\simeq(1+\epsilon)F_\infty\big(x/(1+\epsilon)\big)$ with
$\epsilon(\beta)\to0$. Such a deformation shifts $R$ and $R_c^{\rm B}$
together: its contribution to $\beta\,\partial F/\partial\beta$ is
proportional to $F-xF_x$, which vanishes at $x=R/R_*$, so it does not
move the zero of Eq.~\eqref{eq:dmdA} relative to $R$.

\subsection{Confinement bounds and the unbounded atmosphere}
\label{sec:bounds}

The rigid wall is not merely one admissible choice of outer boundary;
it is the most stabilizing one. Consider any separated, self-adjoint
condition at $R_{\rm out}$, $\zeta(R_{\rm out})\cos\alpha-\kappa^{-1}\eta(R_{\rm out})\sin\alpha=0$,
which in Pr\"ufer form reads $\theta(R_{\rm out})=\alpha+n\pi$ with
$\alpha\in(0,\pi]$; TOVFO and TOVFB correspond to $\alpha=\pi/2$ and
$\alpha=\pi$. Because $\theta(R_{\rm out};\omega^2)$ increases monotonically
with $\omega^2$, the fundamental eigenvalue increases monotonically
with $\alpha$. Its maximum, at $\alpha=\pi$, is the boxed eigenvalue.
Thus, for every such boundary condition,
\begin{equation}
  \omega^2_{0,\alpha}(R_{\rm out})\le\omega^2_{0,\rm B}(R_{\rm out}).
\end{equation}

Moreover, $\omega^2_{0,\rm B}$ decreases monotonically with $R_{\rm out}$:
the boxed eigenfunction on $[0,R_{\rm out}]$, extended by zero, is an
admissible trial function in Eq.~(\ref{eq:rayleigh}) on any larger
domain. The same construction applies to the complete atmosphere. For
$R_{\rm out}>R_c^{\rm B}$, the boxed fundamental mode, extended by zero to
$r>R_{\rm out}$, is a compactly supported perturbation with regular inner
behavior and negative Rayleigh quotient. The radial pulsation operator
on $[0,\infty)$ therefore has negative spectrum for any boundary
condition at infinity whose form domain contains compactly supported
perturbations. Because the trial function vanishes outside $R_{\rm out}$, this
conclusion is insensitive to the self-similar, non-asymptotically flat
structure of the outer atmosphere.

Taken together, these results imply that a TOVF whose fluid extends
beyond $R_c^{\rm B}\simeq R$ is radially unstable, whether it is
truncated by any separated, self-adjoint outer boundary condition or retains
its complete atmosphere. Confinement can stabilize the configuration
only if it acts within the narrow interval
$R_c^{\rm O}<R_{\rm out}<R_c^{\rm B}$, i.e., $0.90\,R\lesssim R_{\rm out}\lesssim R$
for large $\beta$, or at smaller radii. Frequency-dependent boundary
conditions, such as coupling to a dynamical exterior medium, lie
outside this argument.

\section{Conclusions}
\label{sec:conclusions}

We have investigated the stability against radial perturbations
of \acp{tovf}
with a radiation equation of state, $p=\rho/3$.
These configurations consist of a negative point-mass
singularity surrounded by positive-density fluid.
Their mass profiles exhibit a rapid transition from the
negative-mass interior to a Schwarzschild-like plateau,
followed by an extended atmosphere approaching a
self-similar asymptotic solution. Since the pressure does
not vanish at any finite radius, the radial stability
problem requires the specification of an outer boundary
condition.

We have considered two boundary-value problems:
TOVFO, for which the Lagrangian pressure perturbation
vanishes at the outer boundary, $\eta(R_{\rm out})=0$, and TOVFB,
for which the radial displacement vanishes,
$\zeta(R_{\rm out})=0$. Using the Pr\"ufer phase method, we
identify the fundamental radial mode and determine the
critical radius $R_c$ at which its squared frequency
vanishes. For both boundary conditions, the fundamental
mode is stable for $R_{\rm out}<R_c$ and unstable for
$R_{\rm out}>R_c$ in the range examined. For the boxed condition,
this holds for every $R_{\rm out}$, since the fundamental boxed
eigenvalue decreases monotonically with $R_{\rm out}$.

The numerical results reveal that the critical radii
approximately follow the same scaling with the central
negative mass parameter $\mu$ as the equilibrium
firehole radius $R$ and plateau mass $M$.
For families parametrized by the inner lapse
normalization $A=\mu^q/\sqrt{5}$, we find
\begin{equation}
R,\ M,\ R_c \propto \mu^{(1+2q)/5},
\end{equation}
in agreement with the equilibrium scaling discussed
in Sec.~\ref{sec:equilibria}. The dependence on $q$ reflects the choice
of inner normalization used to construct a family of
solutions, rather than a unique mass--radius relation
determined by the TOV equations. A scaling symmetry of the
TOV equations reduces the singular solutions, up to an overall
scale, to a one-parameter family labeled by $\beta=\mu/R_*$;
all dimensionless ratios, including $R_c/R$, depend only on
$\beta$.

A central result of this study is that the two boundary
conditions lead to systematically different critical
radii. For TOVFO, the marginal-stability radius lies
inside the characteristic firehole radius,
$R_c^{\rm O}<R$, approaching $R_c^{\rm O}\simeq0.898\,R$ for
large $\beta$. In contrast, for TOVFB, the critical
radius nearly coincides with it,
$R_c^{\rm B}\simeq R$, over the parameter range
examined. Thus, confinement of the radial displacement
allows the configuration to remain stable over a larger
radial domain, extending nearly to the firehole radius.

The different stability thresholds can be understood
through the variational properties of the radial
pulsation operator. For the same equilibrium background
and outer radius, the rigid-wall condition restricts
the space of admissible perturbations relative to the
open condition. Consequently, the fundamental squared
frequencies satisfy
\begin{equation}
\omega_{0,\rm O}^2(R_{\rm out})
\leq
\omega_{0,\rm B}^2(R_{\rm out}),
\end{equation}
under the corresponding self-adjoint boundary
conditions. This establishes the stabilizing effect
of confinement and is consistent with the ordering
$R_c^{\rm O}<R_c^{\rm B}$.

The zero-frequency eigenfunctions provide a complementary
interpretation. Both boundary-value problems select
different truncations of the same regular zero-mode
solution. The first zero of the momentum-like variable
$\eta=P\zeta'$ determines $R_c^{\rm O}$, while the
subsequent first zero of $\zeta$ determines $R_c^{\rm B}$.
Furthermore, the zero-frequency identity
$\mathcal{N}(r)=\zeta(r)\eta(r)$, in the limit of a
vanishing inner integration radius, relates the cumulative
variational numerator directly to the eigenfunctions.
Its behavior illustrates why the sign of a cumulative
contribution at an intermediate radius cannot by itself
determine the stability of a configuration with a
different outer boundary condition.

This regular zero mode is the displacement that connects
neighboring equilibria with the same central mass, generated
by varying the inner lapse normalization $A$. The boxed
threshold is therefore a turning point: $R_c^{\rm B}$ is the
radius at which the mass enclosed by the wall is stationary
along this family. Combined with the scaling symmetry, this
shows that the rigid-wall zero mode vanishes where the
equilibrium compactness $m/r$ is extremal in the limit of
large $\beta$, which is the definition of the firehole radius.
The near-coincidence $R_c^{\rm B}\simeq R$ is thus explained,
with $R_c^{\rm B}/R-1\simeq 0.009/\beta$.

Our results demonstrate that the radial stability of
TOVFs cannot be characterized independently of the
boundary conditions imposed on their extended fluid
atmosphere. In particular, the instability of an open
configuration does not imply instability of the same
background when confined by a rigid boundary placed in the
interval $R_c^{\rm O}<R_{\rm out}<R_c^{\rm B}$, that is,
approximately between $0.9\,R$ and $R$.
At the same time, the presence of a negative central
singularity does not, by itself, determine the sign of
the fundamental radial eigenvalue. Extending the analysis
to physically specified, dynamical outer boundaries remains
an open question.

An important implication of our results concerns the
possibility of regarding TOVFs as isolated, stable
compact objects. Among all separated, self-adjoint outer
boundary conditions, the rigid wall yields the largest
fundamental eigenvalue, and even under this condition the
critical radius nearly coincides with the characteristic
firehole radius, $R_c^{\rm B}\simeq R$. Extending
the fluid domain beyond this radius introduces
an unstable fundamental radial mode. Moreover, the boxed
unstable mode, extended by zero, is a compactly supported
perturbation of the complete, unbounded atmosphere with
negative Rayleigh quotient. Since TOVFs
possess an atmosphere extending indefinitely, these
configurations are radially unstable as isolated objects,
and stability requires an additional physical mechanism
that confines the surrounding fluid on a scale comparable
to the firehole radius.

\begin{acknowledgments}
This work was partially supported by NSF grants PHY-2411068 and PHY-2207780. The authors thank Fumika Suzuki, Wojciech Zurek, and Natalia Dinkus for helpful discussions. During the preparation of this manuscript, the authors used
Claude (Anthropic), a large language model, to review the draft. The
model independently verified the equilibrium and zero-frequency
pulsation integrations, reproducing the numerical results reported
here, including the fits in Table~\ref{tab:tovf_scaling}, and suggested
revisions to notation and presentation. The authors
take full responsibility for all analytic results,
numerical computations, and conclusions of this paper.
\end{acknowledgments}

%\bibliography{biblio,new_bib_entries}% Produces the bibliography via BibTeX.

\end{document}